\documentclass[twocolumn,nofootinbib,amsmath,amssymb,a4paper]{revtex4}

\usepackage{graphicx}
\usepackage{epsfig}
\usepackage{dcolumn}
\usepackage{bm}
\def\bc{\begin{center}}
\def\ec{\end{center}}
\def\beq{\begin{equation}}
\def\eeq{\end{equation}}
\def\bs{\begin{slide}}
\def\es{\end{slide}}
\newcommand{\bmath}{\begin{displaymath}}
\newcommand{\emath}{\end{displaymath}}
\newcommand{\beqn}{\begin{eqnarray}}
\newcommand{\eeqn}{\end{eqnarray}}
\newcommand{\beqns}{\begin{eqnarray*}}
\newcommand{\eeqns}{\end{eqnarray*}}
\newcommand{\ba}{\begin{array}{c}} 
\newcommand{\bat}{\begin{array}{cc}} 
\newcommand{\ea}{\end{array}}

\newcommand{\gev}{\, \mbox{GeV}}
\newcommand{\mev}{\, \mbox{MeV}}

\newcommand{\fkp}{$f_+^{K^0 \pi^-}(0) \,$}
\begin{document}

\title{Analytical approaches to the calculation of $f_+^{K^0 \pi^-}(0)$
\footnote{Talk given at the 4$^{th}$ International Workshop on the CKM Unitarity
Triangle, CKM2006, 12$^{th}$-16$^{th}$ December 2006, Nagoya (Japan). Report IFIC/07-14.}
}

\author{J. Portol\'es}
 \email{Jorge.Portoles@ific.uv.es}
\affiliation{%
Instituto de F\'{\i}sica Corpuscular, IFIC, \\ CSIC-Universitat de Val\`encia, \\ Apt.~Correus 22085,
E-46071 Val\`encia, Spain
}%

\begin{abstract}
 $K_{\ell 3}$ decays constitute, at present, the golden modes to extract $|V_{us}|$ from
 experimental data. Its incertitude is dominated by the theoretical
 error in the determination of the vector form factor at zero transfer of momentum. I review
 the most relevant analytical approaches for the calculation of this parameter. 
 
\end{abstract}

\maketitle

\section{Introduction}
Semileptonic kaon decays have the potential to provide information on the $V_{us}$ CKM matrix element. 
In principle $K_{\ell 2}$ decays offer the most simple setting, but the fact that these are driven by the axial-vector
QCD current, hence the kaon decay constant, diminish their capability~: $F_K$ is not given by symmetry and 
has to be input from theory, e.g. lattice. Something analogous happens in $K_{\ell 3}$ processes where 
another parameter guided by strong interactions, i.e. the vector form factor at zero transfer of momentum, has to 
be determined. However several fortunate features happen together in the later decays that convert them in an
excellent system to extract information on $|V_{us}|$~:
i) Only the QCD vector current contributes to $K_{\ell 3}$ decays. Due to the conservation of this current
in the $SU(N)_F$ limit (CVC), the normalization at zero transfer of momentum is known in this limit;
ii) Corrections to the above result only start at second order in the symmetry breaking parameter, i.e.
${\cal O}\left[\left(m_s - m_d\right)^2\right]$ for three flavours
\cite{Ademollo:1964sr}.
\par
The form factors of the vector current in $K_{\ell 3}$ decays are defined by~:
\begin{eqnarray} \label{eq:fff}
\langle \pi^-(p_{\pi}) | \overline{s} \gamma_{\mu} u | K^0(p_K) \rangle  & =   & \\ 
f_{+}^{K^0 \pi^-}(t) \left[ \left( p_K +p_{\pi} \right)_{\mu}   - \frac{\Delta_{K \pi}}{t} q_{\mu} \right] & + &   
f_{0}^{K^0 \pi^-}(t) \frac{\Delta_{K \pi}}{t} q_{\mu} \nonumber ,
\end{eqnarray}
where $q_{\mu} = (p_K - p_{\pi})_{\mu}$, $t=q^2$ and $\Delta_{K \pi} = M_K^2 - M_{\pi}^2$. The functions 
$f_+^{K^0 \pi^-}(t)$ and
$f_0^{K^0 \pi^-}(t)$ are known as the vector and scalar form factors, respectively. The fact that
$f_{+}^{K^0 \pi^-}(0) = f_{0}^{K^0 \pi^-}(0)$ allows us to write 
$f_{+,0}^{K^0 \pi^-}(t) = f_{+}^{K^0 \pi^-}(0) \left[ 1 + \lambda_{+,0}' \frac{t}{M_{\pi^+}^2} + ... \right]$, and
then factorize $f_+^{K^0 \pi^-}(0)$ in the partial width as~:
\begin{eqnarray} \label{eq:gamma}
\Gamma\left( K_{\ell 3 [\gamma]}^0 \right) & \sim & \, |\, V_{us} \, f_+^{K^0 \pi^-}(0) \, |^2 \, 
I^{K \ell} \left( \lambda_{+,0}'. ... \right) \\
& & \times  \,  \left[ 1 + \Delta_{SU(2)} + \Delta_{EM} \right] \, , \nonumber
\end{eqnarray}
where $\Delta_{SU(2)}$ and $\Delta_{EM}$ embody isospin breaking effects of strong and electromagnetic origin,
respectively \cite{Cirigliano:2001mk}. As can be seen from Eq.~(\ref{eq:gamma}), the determination of $|V_{us}|$ from
$K_{\ell 3}$ decays relies essentially in our ability to ascertain an accurate value for \fkp.
\par 
We have some basic model-independent knowledge on \fkp. In the $SU(3)$ limit, CVC establishes that
$f_{+}^{K^0 \pi^-}(0) |_{SU(3)} = 1$. Moreover, a known sum rule \cite{DeAlfaro:1969rf} asserts that 
$f_{+}^{K^0 \pi^-}(0) < 1$ and, consequently, flavour breaking corrections given by the Ademollo-Gatto 
theorem \cite{Ademollo:1964sr} should be negative. Given this setting our goal in order to provide a close
determination of \fkp \ is to evaluate $SU(3)$ breaking corrections to $K_{\ell 3}$ decays. The appropriate
framework to perform this task is Chiral Perturbation Theory ($\chi$PT) \cite{Weinberg:1978kz,Gasser:1983yg}.
 
\section{\fkp \ up to ${\cal O}(p^6)$ in $\chi$PT}
$SU(3)$ breaking contributions span the deviation of \fkp \ from unity~:
\begin{equation} \label{eq:chiralex}
f_{+}^{K^0 \pi^-}(0) \, - \, 1 \, = \, f_{p^4} \, + \, f_{p^6} \, + ... \, ,
\end{equation}
where $f_{p^n}$ is of ${\cal O}(p^n)$ in the chiral expansion. The ${\cal O}(p^4)$ correction amounts to a pure
one-loop calculation (there is no local contribution from counterterms) and then it is essentially free from 
uncertainties, giving $f_{p^4} = -0.0227$ \cite{Leutwyler:1984je,Gasser:1984ux}. At ${\cal O}(p^6)$ the situation
is more complex. The first estimate was obtained by relating \fkp \ to the matrix element of the vector charge
between a kaon and a pion in the infinite-momentum limit \cite{Leutwyler:1984je}. This procedure understates 
chiral log contributions and gives~:
\begin{equation} \label{eq:lr}
f_{p^6}^{LR} \, = \, -0.016 (8) \, ,
\end{equation}
that essentially coincides with the result provided by the light-front quark model \cite{Choi:1998jd}.
\par
Within $\chi$PT all kind of possible contributions appear \cite{Post:2001si,Bijnens:2003uy}~:
\begin{equation} \label{eq:op6loop}
f_{p^6} \, = \, f_{p^6}^{2-\mathrm{loops}}\left( \mu \right) \, + \, 
f_{p^6}^{L_i \times \mathrm{loop}}\left( \mu \right) \, + \, f_{p^6}^{\mathrm{tree}}\left( \mu \right) \; .
\end{equation}
Loop terms can be evaluated with small uncertainty \cite{Bijnens:2003uy} obtaining~:
\begin{eqnarray}
f_{p^6}^{2-\mathrm{loops}}\left( M_{\rho} \right) & = & 0.0113 \; , \nonumber \\ 
f_{p^6}^{L_i \times \mathrm{loop}}\left( M_{\rho} \right) & = & -0.0020 (5) \; , 
\end{eqnarray}
while the tree--level part is, a priori, unknown and can be expressed in terms of chiral low-energy couplings
(LECs) of the chiral Lagrangian both at ${\cal O}(p^4)$ \cite{Gasser:1984gg} and 
${\cal O}(p^6)$ \cite{Bijnens:1999sh}~:
\begin{eqnarray} \label{eq:fp6tree}
f_{p^6}^{\mathrm{tree}}\left( M_{\rho} \right) & = & 8 \, \frac{\Delta_{K \pi}^2}{F_{\pi}^2} \, 
\times  \\ & & 
\left[ \frac{\left(L_5^r\left(M_{\rho}\right)\right)^2}{F_{\pi}^2} - C_{12}^r\left(M_{\rho}\right) - 
C_{34}^r\left(M_{\rho}\right) \, \right] \, , \nonumber
\end{eqnarray}
where $F_{\pi}= 92.4 \mev$ is the pion decay constant.
While the $L_i$ couplings are rather well known both from direct phenomenology \cite{Amoros:2001cp} and 
on more theoretical grounds \cite{Ecker:1988te}, our knowledge on the ${\cal O}(p^6)$ $C_i$ LECs is still 
rather poor. As a consequence
it is difficult to provide a determination for the ${\cal O}(p^6)$ contribution to \fkp . 
In Ref.~\cite{Bijnens:2003uy} the authors did combine their loop evaluation together with the Leutwyler \& Roos (LR)
result $f_{p^6}^{LR}$ in Eq.~(\ref{eq:lr}), assuming that the later stands for the $C_i$ contribution. As suggested
in Ref.~\cite{Cirigliano:2005xn} it is more natural to presume that $f_{p^6}^{LR}$
indicates the full tree contribution $f_{p^6}^{\mathrm{tree}}$. This would give~:
\begin{equation}
f_{p^6}^{BT} \, = \, -0.007 (9) \, . 
\end{equation}
Several strategies to determine $f_{p^6}$ have been
developed~: on one side lattice quenched and unquenched evaluations are now at hand \cite{zanotti}, on the other
there are theoretical tools that may provide a determination of the unknown LECs. The latter are usually encoded as
analytical evaluations and will be considered in the following.

\section{LECs from the scalar $K \pi$ form factor}
The ${\cal O}(p^6)$ determination of the $K_{\ell 3}$ form factors shows that the scalar form factor in 
Eq.~(\ref{eq:fff}) depends on the same ${\cal O}(p^6)$ LECs than $f_{p^6}^{\mathrm{tree}}$ 
\cite{Bijnens:2003uy,Jamin:2004re}. It can be written as~:
\begin{eqnarray}
f_0^{K^0 \pi^-}(t) & = & f_{+}^{K^0 \pi^-}(0) + \overline{\Delta}(t) + \left( \frac{F_K}{F_{\pi}} -1 \right) 
\frac{t}{\Delta_{K \pi}} \nonumber \\
& & \! \! \! + \, \frac{8}{F_{\pi}^2} \left[ \Sigma_{K \pi} \left( 2 C_{12}^r + C_{34}^r \right) t - 
C_{12}^r t^2 \right] , 
\end{eqnarray}
where $\Sigma_{K \pi}= M_K^2+M_{\pi}^2$, $\overline{\Delta}(t)$ receives contributions from ${\cal O}(p^4)$
and ${\cal O}(p^6)$ and its only dependence in LECs involves $L_i$ at ${\cal O}(p^6)$. 
Moreover $\overline{\Delta}(0)=0$. Hence it is possible to extract the $C_i$ LECs from data if a good knowledge on the
slope and curvature of the scalar form factor is available~:
\begin{eqnarray} \label{eq:slocur}
2 C_{12}^r + C_{34}^r & \sim & \frac{d}{dt} f_0^{K^0 \pi^-}(t) \Big|_{t=0}  \, , \nonumber \\
C_{12}^r & \sim & \frac{d^2}{dt^2} f_0^{K^0 \pi^-}(t) \Big|_{t=0}  \, . 
\end{eqnarray}
This procedure has been addressed in Ref.~\cite{Jamin:2004re}. The $K \pi$ scalar form factor was reconstructed
from a coupled-channel dispersion relation analysis from $K \pi$ scattering data \cite{Jamin:2001zq} and then
employed to obtain information on the LECs using Eq.~(\ref{eq:slocur}) above. By adding 
$f_{p^6}^{\mathrm{tree}}(M_{\rho})$ to the chiral logs they obtain~:
\begin{equation}
f_{p^6}^{\mathrm{JOP}} \, = \, -0.003 (11) \; ,
\end{equation}
where most of the error arises from the determination of the $C_i$ LECs. This tiny value comes from an almost
complete cancellation between the loop and tree contributions. 
\par
The determination of the $K \pi$ scalar form factor from $K \pi$ scattering data shows that, in order to 
solve the coupled channel dispersion relations, two integration constants are required. The authors
of Ref.~ \cite{Jamin:2001zq} have input $f_0^{K^0 \pi^-}(0)$ and $f_0^{K^0 \pi^-}\left( \Delta_{K \pi} \right)$. 
Accordingly
their evaluation can be viewed more as a consistency check of the fact that 
$f_{+}^{K^0 \pi^-}(0) =f_0^{K^0 \pi^-}(0)$ than a pure determination of the $C_i$ LECs.

\section{LECs from resonance saturation}
A systematic procedure to account for the resonance contributions to chiral LECs involves the construction of a 
Lagrangian theory in terms of $SU(3)$ (pseudo-Goldstone mesons) and $U(3)$ (heavier resonances) flavour multiplets
as active degrees of freedom. This is called Resonance Chiral Theory (R$\chi$T). 
The method was applied in Ref.~\cite{Ecker:1988te} to evaluate the resonance 
contributions to ${\cal O}(p^4)$ chiral LECs and has recently been extended \cite{Cirigliano:2006hb} to look
upon the ${\cal O}(p^6)$ LECs. The scheme relies in two key premises~: i) The most general possible Lagrangian, 
including all terms consistent with assumed symmetry principles provides the most general possible S-matrix
amplitude consistent with analyticity, perturbative unitarity, cluster decomposition and the specified principles
of symmetry \cite{Weinberg:1978kz}; ii) It has been suggested that large-$N_C$ QCD, embodied in an effective local
Lagrangian for the study of meson dynamics, shows features that resemble, both qualitatively and quantitatively, 
the $N_C=3$ case \cite{'tHooft:1973jz}.
\par
The procedure that has been devised amounts to construct the Lagrangian theory by imposing the chiral symmetry
constraints on the pseudo-Goldstone mesons and unitary symmetry on the resonance fields. This guarantees the proper
chiral limit. In addition short-distance dynamics is enforced on the couplings of the theory by demanding 
either the
appropriate asymptotic behaviour for Green functions that are order parameters of the chiral symmetry breaking 
or the high-energy smoothing of form factors of QCD currents \cite{Peris:1998nj,Knecht:2001xc}. 
\par
It is well known that the phenomenological value of ${\cal O}(p^4)$ LECs is saturated by the lightest spectrum
not included in $\chi$PT, i.e. resonances. This result seems to persist at one-loop in 
R$\chi$T \cite{Portoles:2006nr}.
An analogous result for the ${\cal O}(p^6)$ LECs
has still not been reached though there is no hint that indicates it could happen otherwise~: 
resonance saturation seems to be an underlying property stemming from non-perturbative QCD.
\par
\begin{figure}[!t]
\centering

\begin{picture}(300,175)  
\put(105,65){\makebox(50,50){\epsfig{figure=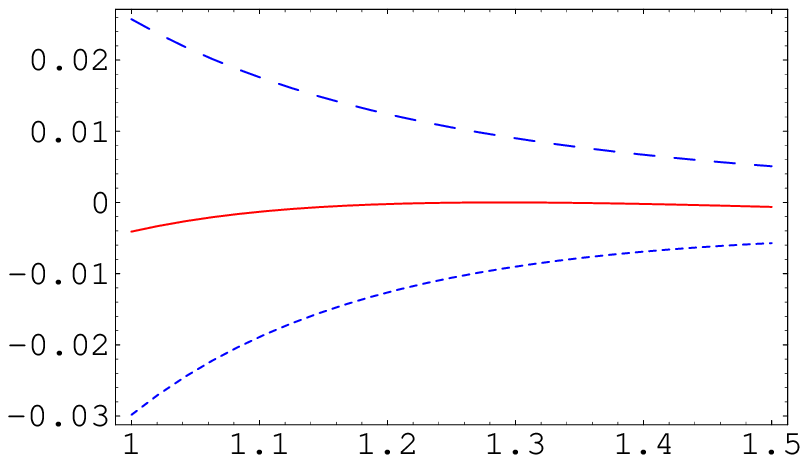,width=8.1cm}}}
\put(190,0){
{\large 
$M_S \ ({\rm GeV})$ 
}
}
\put(2,165){
{\large 
$ f_{p^6}^{\rm tree} (M_\rho)  $ 
}
}
\put(130,130){
{\small
$L_5 \times L_5 / F_\pi^2 $
}
}
\put(130,60){
{\small
$- (C_{12} + C_{34})$
}
}
\end{picture}
\caption{
\label{fig:ftree}
$f_{p^6}^{\rm tree} (M_\rho)$ as a function of $M_S$ for $M_P=1.3$ GeV 
(solid line). 
We also plot the two components: the dashed line represents  the 
term proportional to $L_5 \times L_5$, while the 
dotted line represents the term proportional to 
$- (C_{12} + C_{34})$. 
}
\end{figure}

The analysis, along the lines outlined above, of the $\langle SPP \rangle$ Green function give us the following
contributions of scalar and pseudoscalar resonances to the relevant LECs in Eq.~(\ref{eq:fp6tree})
\cite{Cirigliano:2005xn}~:
\begin{eqnarray}
L_5^{SP} & = & \frac{F_{\pi}^2}{4 M_S^2} \; , \nonumber \\
C_{12}^{SP} & = & - \, \frac{F_{\pi}^2}{8 M_S^4} \; , \\
C_{34}^{SP} & = & \frac{3 F_{\pi}^2}{16 M_S^4} \, + \, \frac{F_{\pi}^2}{16} \left( \frac{1}{M_S^2} - \frac{1}{M_P^2} 
\right)^2 \; , \nonumber
\end{eqnarray}
where $M_S$ ($M_P$) is the mass of the lightest multiplet of scalar (pseudoscalar) resonances. Substituting these
values into $f_{p^6}^{\mathrm{tree}}$ in Eq.~(\ref{eq:fp6tree}) we get~:
\begin{equation}
f_{p^6}^{\mathrm{tree}}(M_{\rho}) \, = \, - \, \frac{\Delta_{K \pi}^2}{2 M_S^4} \, \left( 1 - \frac{M_S^2}{M_P^2}
\right)^2 \; .
\end{equation}
As can be seen in Figure~\ref{fig:ftree} there is an almost complete cancellation between the two different
counterterm contributions. This is a fortunate feature because the resulting dependence in the scalar resonance
mass happens to be tiny. For $M_P = 1.3  \gev$ and adding the chiral logs we get
\footnote{For a detailed discussion on the error
of this prediction please resort to Ref.~\cite{Cirigliano:2005xn}.}~:
\begin{equation}
f_{p^6}^{\mathrm{res}} \, = \, 0.007 (12) \; .
\end{equation}
Notice that the sign of the central value is opposite to previous predictions above. However this is not significant
because, as in $f_{p^6}^{BT}$ or $f_{p^6}^{JOP}$, the result is compatible with zero. 

\section{Comparison}
Lattice results, both quenched \cite{Becirevic:2004ya} or unquenched  \cite{zanotti,Dawson:2006qc}
are available. In Table~\ref{tab:uno} we compare these with the results from 
analytical determinations reviewed above. Two trends are easily revealed~: i) Lattice results, both 
quenched and unquenched, are in excellent agreement with the result by LR; ii) Analytical
determinations driven by chiral logs and evaluations of the ${\cal O}(p^6)$ chiral LECs show a clear
tension with the lattice ones, namely when LECs are determined by resonance saturation. 
\par
\begin{table}
\caption{\label{tab:uno} Comparison of different predictions for $f_{+}^{K^0 \pi^-}(0)$.}
\begin{center}
\begin{ruledtabular}
\begin{tabular}{ll}
\multicolumn{1}{c}{Reference} &
\multicolumn{1}{c}{$f_{+}^{K^0 \pi^-}(0)$} \\
\hline
Leutwyler \& Roos (LR) \cite{Leutwyler:1984je} &  $0.961(8)$ \\
Lattice (quenched) \cite{Becirevic:2004ya} & $0.960(9)$ \\
Lattice (unquenched) \cite{Dawson:2006qc}& $0.968(11)$ \\
Lattice (unquenched) \cite{zanotti}  & $0.961(5)$ \\
$\chi$-logs +  LR ($\chi_{LR}$) \cite{Bijnens:2003uy} & $0.971(9)$ \\
$\chi$-logs + $f_{p^6}^{\mathrm{tree}}$ [$K \pi$ scalar f.f.] \cite{Jamin:2004re} & $0.974(11)$ \\
$\chi$-logs + $f_{p^6}^{\mathrm{tree}}$ [resonance saturation] \cite{Cirigliano:2005xn} & $0.984(12)$ \\
\end{tabular} 
\end{ruledtabular}
\end{center}
\end{table}
The LR prediction plays down ${\cal O}(p^6)$ chiral logs hence it may 
be an accurate prescription as long as these are not important. However the chiral determination (\ref{eq:op6loop})
shows that part of them are indeed large and positive, hence
the addition of chiral logs and LR produces a cancellation that manifests in the $\chi_{LR}$
result of Table~\ref{tab:uno}. If the LR prediction stands for the local ${\cal O}(p^6)$ contribution it can
be compared with $f_{p^6}^{\mathrm{tree}}(M_{\rho})$ showing a clear discrepancy~: all LEC determinations 
conclude that the final result for $f_{p^6}^{\mathrm{tree}}(M_{\rho})$ is tiny (consistent with zero) while
$f_{p^6}^{LR}$ is comparatively large. 
\par
There is also a point to be understood with the lattice analyses. The latter unquenched result 
\cite{zanotti} agrees fully with the quenched determination from Ref.~\cite{Becirevic:2004ya}. This could be
expected if chiral logs were definitely small but Eq.~(\ref{eq:op6loop}) shows that this is not the case. 
The unquenched determination could miss the curvature provided by the chiral logs in the necessary 
extrapolation performed to reach the physical values of the masses and then its tiny error  should
be indeed larger.
Finally there is another recourse. Lattice evaluations determine $\Delta f = f_{+}^{K^0 \pi^-}(0) - 1 - f_{p^4}$,
i.e. they cover all chiral $SU(3)$ symmetry breaking corrections except the first one. Though
it would be most rare, if chiral corrections higher than ${\cal O}(p^6)$ were large these could bring an 
explanation to the discrepancy between the chiral and lattice results. However at present
there is no expectation for an estimate of chiral ${\cal O}(p^8)$ contributions.
\par
Determinations of  $|V_{us}|$ from $K_{\ell 3}$ using the values of \fkp \ in Table~\ref{tab:uno} 
could also have a say to discern in the discussion just sketched. Taking into account the later published value
$f_{+}^{K^0 \pi^-}(0) \, |V_{us}| = 0.2164 (4)$ \cite{Versaci:2007nx}, the {\em LR} value gives
$|V_{us}| = 0.2252 (19)$, {\em $\chi_{LR}$} lowers it to  $|V_{us}| = 0.2229 (21)$ while {\em resonance saturation} 
brings it down to $|V_{us}| = 0.2199(27)$. All of them are away from the value obtained by the unitarity
constraint $|V_{us}^{\mathrm{unitarity}}|=0.2275 (12)$ \cite{Yao:2006px} though, within errors, the latter is 
compatible with the {\em LR} evaluation. 

\section{Outlook}
The study of \fkp \ is an all-important key to obtain an accurate value of $|V_{us}|$ from $K_{\ell 3}$ decays.
Driven by $SU(3)$ breaking contributions its determination is, however, tampered by the non-perturbative
nature of these corrections. 
As we have seen, and though the errors are still large, there is essentially no
agreement between lattice evaluations and LR on one side, and chiral evaluations on the other. Several points
need to be asserted on the ${\cal O}(p^6)$ contributions~: i) The size of the chiral logs has to be confirmed. 
If they stand as at
present then unquenched lattice results should revise their extrapolation and error estimate. In addition the
LR prediction should also be amended; ii) Local
contributions have also to be analysed. Though all LEC determinations agree that their contribution is tiny, this
conclusion is at odds with the LR evaluation. 
\par
In conclusion more work is needed in order to disentangle the ${\cal O}(p^6)$ contributions to \fkp. 
A check of the
size of the chiral logs would be very much welcome and, moreover, whether there is a failure of the resonance
saturation hypothesis, employed in the determination of ${\cal O}(p^6)$ chiral LECs, or of the other models 
appointed to evaluate
these local contributions, the situation deserves closer attention.

\begin{acknowledgments}
\vspace*{-0.3cm}
This work has been supported in part by the EU
MRTN-CT-2006-035482 (FLAVIAnet), by MEC (Spain) under grant
FPA2004-00996 and by Generalitat Valenciana under grants ACOMP06/098 and GV05/015.
\end{acknowledgments}

\end{document}